\documentclass[3p, sort&compress]{elsarticle}
\usepackage[colorlinks=true, allcolors=blue]{hyperref}
\usepackage{amsmath}
\usepackage{amssymb}
\usepackage{amsfonts}
\usepackage{mathrsfs}
\usepackage{bm}
\newcommand{\lt}{\left}
\newcommand{\rt}{\right}
\newcommand{\pa}{\partial}
\newcommand{\ve}{\varepsilon}
\newcommand{\Ha}{\mathcal{H}}
\newcommand{\br}{\mathbf{r}}
\newcommand{\bx}{\mathbf{x}}
\newcommand{\bh}{\mathbf{h}}
\newcommand{\bq}{\mathbf{q}}
\newcommand{\hq}{\hat{\mathbf{q}}}
\newcommand{\bk}{\mathbf{k}}
\newcommand{\bu}{\mathbf{u}}
\begin{document}
\begin{frontmatter}

\title{Scaling behavior of crystalline membranes: an $\ve$-expansion approach}

\author[mainaddress]{Achille Mauri\corref{correspondingauthor}}
\cortext[correspondingauthor]{Corresponding author}
\ead{a.mauri@science.ru.nl}
\author[mainaddress]{Mikhail I. Katsnelson}
\ead{m.katsnelson@science.ru.nl}
\address[mainaddress]{Radboud University, Institute for Molecules and Materials, Heyendaalseweg 135, 6525 AJ Nijmegen, The Netherlands}

\begin{abstract}
We study the scaling behavior of two-dimensional (2D) crystalline membranes in the flat phase by a renormalization group (RG) method and an $\ve$-expansion. Generalization of the problem to non-integer dimensions, necessary to control the $\ve$-expansion, is achieved by dimensional continuation of a well-known effective theory describing out-of-plane fluctuations coupled to phonon-mediated interactions via a scalar composite field, equivalent for small deformations to the local Gaussian curvature. The effective theory, which will be referred to as  Gaussian curvature interaction (GCI) model, is equivalent to theories of elastic $D$-dimensional manifolds fluctuating in a $(D + d_{c})$-dimensional embedding space in the physical case $D = 2$ for arbitrary $d_{c}$. For $D\neq 2$, instead, the GCI model is not equivalent to a direct dimensional continuation of elastic membrane theory and it defines an alternative generalization to generic internal dimension $D$. After decoupling interactions through a Hubbard-Stratonovich transformation, we study the GCI model by perturbative field-theoretic RG within the framework of an expansion in $\ve = (4 - D)$. We calculate explicitly RG functions at two-loop order and determine the exponent $\eta$ characterizing the long-wavelength scaling of correlation functions to order $\ve^{2}$. The value of $\eta$ at this order is shown to be insensitive to Feynman diagrams involving vertex corrections. As a consequence, the self-consistent screening approximation for the GCI model is shown to be exact to O($\ve^{2}$). In the physical case of a single out-of-plane displacement field, $d_{c} = 1$, the O($\ve^{2}$) correction is suppressed by a small numerical prefactor. As a result, despite the large value of $\ve = 2$,  extrapolation of the first and second order results to $D = 2$ leads to very close numbers, $\eta = 0.8$ and $\eta \simeq 0.795$. The calculated exponent values are close to earlier reference results obtained by non-perturbative renormalization group, the self-consistent screening approximation and numerical simulations. These indications suggest that a perturbative analysis of the GCI model could provide an useful framework for accurate quantitative predictions of the scaling exponent even at $D = 2$.
\end{abstract}
\end{frontmatter}

\section{Introduction}
The statistical properties of two-dimensional (2D) membranes fluctuating in a higher-dimensional space have been the subject of extensive investigations~\cite{nelson_statistical}. Models of flexible surfaces are relevant in several contexts, from string theories and quantum gravity to biological and condensed matter systems. In recent years, the experimental isolation of graphene and other atomically-thin two-dimensional materials has stimulated vast interest in the behavior of solid membranes, characterized by an internal crystalline order and fixed connectivity between consituent atoms~\cite{katsnelson_graphene, katsnelson_acr_2013}. The statistical mechanical properties of membranes of this class have been studied thoroughly for decades, in connection with polymerized layers and biological membranes.

A crucial prediction in the theory of fluctuating crystalline membranes is the existence of a thermodynamically stable flat phase at low temperatures~\cite{nelson_jpf_1987}. In this phase rotational symmetry is spontaneously broken, the state of the system is macroscopically planar and the vectors normal to the surface exhibit long-range orientational order. Rotational invariance is  restored at a finite-temperature transition, above which the system behaves as a crumpled manifold~\cite{nelson_statistical, paczuski_prl_1988}. The phase structure already indicates that the behavior of elastic membranes is very peculiar~\cite{aronovitz_jpf_1989, guitter_jpf_1989}. Under broad conditions, the Mermin-Wagner theorem forbids spontaneous breakdown of continuous symmetries in two dimensions and crystalline membranes seem to violate this rule~\cite{nelson_statistical}. The reason can be traced to the existence of long-range interactions mediated by the  elasticity of the surface, a consequence of its internal crystalline order~\cite{nelson_jpf_1987, guitter_jpf_1989}. Transverse fluctuations, which tend to fold the membrane in the third dimension, are strongly suppressed by their coupling with shear deformation. Interactions between out-of-plane undulations and 'massless' in-plane phonons lead to power-law renormalizations and non-Gaussian scaling behavior of fluctuations, stabilizing the flat phase~\cite{nelson_statistical, nelson_jpf_1987, aronovitz_prl_1988, aronovitz_jpf_1989, guitter_jpf_1989}.

The quantitative description of scaling behavior in flat membranes has been addressed by several field theoretic approaches. A systematic analysis can be developed by considering membranes embedded in a space of large dimensionality. Models for $D$-dimensional membranes in a $d$-dimensional space were solved up to order $1/d_{c}$ in an expansion for large codimension $d_{c} = (d- D)$\footnote{In Ref.~\cite{david_epl_1988}, the exponent for large embedding dimension is reported as $2/d$. The expressions $2/d$ and $2/d_{c}= 2/(d-D)$ are equal to leading order for large dimension.}\cite{david_epl_1988, aronovitz_jpf_1989, guitter_prl_1988, guitter_jpf_1989, gornyi_prb_2015}. The solution confirms stability of a flat phase for $D = 2$ and scaling of correlation functions. The effective bending rigidity of the system was found to diverge in the long-wavelength limit as $\kappa(q) \approx q^{-\eta}$, with $\eta = 2/d_{c} + $O$(1/d_{c}^{2})$. In a recent work, the large-$d_{c}$ expansion has been extended to second order, leading to $\eta = 2/d_{c}-(73-68\zeta(3))/(27d_{c}^{2})+.. \simeq 2/d_{c}-0.32/d_{c}^{2} + {\rm O}(1/d_{c}^{3})$~\cite{saykin_arxiv_2020}.

Perturbative approaches to the scaling behavior were also developed~\cite{aronovitz_prl_1988, guitter_jpf_1989, guitter_prl_1988, bowick_pr_2001}. As it was shown in Ref.~\cite{aronovitz_prl_1988}, scaling exponents can be addressed by a controlled $\ve$-expansion, where $\ve = 4 - D$. The long-wavelength behavior, in this framework, can be determined by a field-theory renormalization group (RG) approach, based on a perturbative expansion of renormalization constants. 
Rotational invariance and the corresponding Ward identities give crucial constraints to the renormalization, reducing the number of independent divergences~\cite{aronovitz_prl_1988, guitter_jpf_1989}. The exponent $\eta$ was determined at leading order in $\ve$ by a one-loop calculation~\cite{aronovitz_prl_1988}:$
\eta = \ve/(2 + d_{c}/12)$. 
An important consequence of rotational symmetry has been derived in the perturbative RG framework. While the bending rigidity is infinitely strengthened at long-wavelengths, the effective compression and shear moduli are infinitely softened and they scale algebraically as $B(q), \mu(q) \approx q^{\eta_{u}}$. Ward identities impose an exact relation between exponents: $\eta_{u} = 4 - D - 2 \eta$~\cite{aronovitz_prl_1988, guitter_jpf_1989}. The large-$d_{c}$ and the $\ve$-expansion methods have been extended to describe the response of membranes to a weak external tension. Universal laws, characterized by anomalous scaling behavior have been predicted~\cite{guitter_prl_1988, guitter_jpf_1989, burmistrov_prb_2018, burmistrov_aop_2018, gornyi_2dmater_2017, saykin_arxiv_2020, kosmrlj_prb_2016}.

The limit of large embedding dimension and the $\ve$-expansion reveal important qualitative features of the flat phase of crystalline membranes. However, it is difficult to achieve quantitative accuracy in the physical case, because neither $\ve  = 2$ nor $1/d_{c} =1$ are small parameters. An improved precision in the calculation of the exponent $\eta$ has been obtained within the framework of the self-consistent screening approximation (SCSA). In this approach, the large-$d_{c}$ expansion is promoted to a closed set of truncated Dyson equations, which can be explicitly solved by power-law Green's functions in the long-wavelength limit~\cite{le-doussal_prl_1992, le-doussal_aop_2018, gazit_pre_2009}. By construction, the SCSA recovers the large-$d_{c}$ result to leading order in $1/d_{c}$. As it was shown in Ref.~\cite{le-doussal_prl_1992}, it also turns out to be exact to first order in $\ve$ as a consequence of the Ward identities of rotational symmetry. For physical membranes, the scaling exponent was calculated at leading and next-to-leading order, giving $\eta \simeq$ 0.821 and $\eta \simeq$ 0.789 respectively~\cite{le-doussal_prl_1992, le-doussal_aop_2018, gazit_pre_2009}. These results are compatible with predictions obtained by numerical simulations~\cite{le-doussal_aop_2018, bowick_pr_2001, gompper_cis_1997, gompper_jpcm_1997, los_prb_2009, los_prb_2017,  troster_prb_2013, troster_pre_2015, gao_jmps_2014, wei_jcp_2014, hasik_prb_2018}.

Recently, the statistical mechanics of crystalline membranes has been revisited by non-perturbative renormalization group (NPRG) techniques~\cite{kownacki_pre_2009, braghin_prb_2010, hasselmann_pre_2011, troster_pre_2015, essafi_prl_2011, coquand_pre_2018}. These approaches allowed to construct global theories for the flat phase and the crumpling transition and to describe scaling behavior at arbitrary dimensions. For physical membranes, analytical calculations lead to $\eta \simeq$ 0.85 with a good agreement between different approximate implementations of the NPRG~\cite{kownacki_pre_2009, braghin_prb_2010, hasselmann_pre_2011}. A numerical Fourier Monte Carlo renormalization group approach lead to a value of $\eta$ of approximately $0.79$~\cite{troster_pre_2015}. The small scatter between NPRG and SCSA results gives a further indication of the accuracy of the self-consistent screening approximation as a method for the quantitative calculation of scaling exponents.

In several approaches to crystalline membranes~\cite{nelson_jpf_1987, aronovitz_jpf_1989, gornyi_prb_2015, gornyi_2dmater_2017, saykin_arxiv_2020, burmistrov_aop_2018, le-doussal_prl_1992, le-doussal_aop_2018, gazit_pre_2009, kosmrlj_prb_2016} the starting point consists in the elimination of in-plane displacements in favor of an effective theory describing out-of-plane fluctuations with long-range phonon-mediated interactions. After neglection of nonlinearities which are irrelevant near the upper critical dimension $D = 4$~\cite{aronovitz_prl_1988}, the elasticity theory of $D$-dimensional crytsalline membranes is quadratic in the in-plane displacements and the effective non-local theory for out-of-plane fluctuations can be explicitly constructed by Gaussian integration. The form of the mediated long-range interactions has a particularly simple form in the physical case $D = 2$. It reduces to a coupling between Gaussian curvatures of the membrane~\cite{nelson_jpf_1987}. For general dimension $D$, instead, an additional independent non-local coupling with tensor structure is generated when in-plane displacements are integrated out~\cite{aronovitz_jpf_1989, le-doussal_prl_1992, le-doussal_aop_2018, gornyi_prb_2015}.

In Refs.~\cite{le-doussal_prl_1992, le-doussal_aop_2018} it was recognized that the effective theory for transverse fluctuations would present, after Hubbard-Stratonovich decoupling of interactions, a simple renormalization structure characterized by the absence of vertex renormalization. This property, which was traced to rotational invariance in the embedding space, lies at the origin of the exactness of the self-consistent screening approximation to O$(\ve)$ in the $\ve$-expansion. In Refs.~\cite{kosmrlj_prb_2016, kosmrlj_prx_2017} a renormalization group approach acting directly on the effective theory for out-of-plane fluctuations was developed. By a momentum-shell technique, performed at $D = 2$, recursion relations were obtained in a one-loop approximation, leading to the prediction $\eta = 4/5$ for flat membranes~\cite{kosmrlj_prb_2016}.

In this work we explore a further RG approach to the flat phase of crystalline membranes. The approach aims at combining the geometric simplicity of the $D = 2$ effective theory for out-of-plane fluctuations (which will be referred to as 'Gaussian curvature interaction', GCI,  model) with the powerful methods of perturbative field-theoretic renormalization. After decoupling interactions through a Hubbard-Stratonovich transformation and continuation to non-integer dimensions, we study the renormalization of the GCI model within dimensional regularization and minimal subtraction.

The renormalized GCI model provides a framework for a systematic calculation of the scaling exponent $\eta$ within an $\ve = (4-D)$-expansion.  By an explicit two-loop calculation, we determine the scaling exponent $\eta$ at order $\ve^{2}$. The results reveal  interesting features. The self-consistent screening approximation for the model turns out to be exact not only at first, but also at second order in $\ve$. Extrapolating the value of the exponent to the physical dimensionality gives $\eta = 0.8$ and $\eta \simeq 0.795$ in the leading and next-to-leading orders. Even if $\ve = 2$ is not much smaller than unity, the $\ve$-expansion appears to be stable at this level. The numerical value of the exponent is close to reference values from NPRG calculations, numerical simulations and the SCSA. 

\section{Model}
Configurations of a crystalline membrane are specified by assigning the coordinates $\br(\bx)$ in the $d$-dimensional embedding space of mass points of the surface, labeled by an internal $D$-dimensional coordinate $\bx$. It is generally assumed that fluctuations in the flat phase are controlled by elasticity and bending rigidity, which penalize deformations of the metric and extrinsic curvature of the surface. A general model consists in the energy functional~\cite{paczuski_prl_1988, aronovitz_jpf_1989, guitter_jpf_1989, aronovitz_prl_1988, david_epl_1988, guitter_prl_1988, gornyi_prb_2015}:
\begin{equation}\label{def:H}
H = \frac{1}{2} \int {\rm d}^{D}x \,\big[\kappa (\pa^{2}\br)^{2} + \lambda U_{\alpha \alpha}^{2} + 2 \mu U_{\alpha \beta} U_{\alpha \beta}\big]~,
\end{equation}
where:
\begin{equation}
U_{\alpha \beta} = \frac{1}{2} (\pa_{\alpha} \br \cdot \pa_{\beta} \br - \delta_{\alpha \beta})
\end{equation}
is the strain tensor, proportional to the deviation of the metric $g_{\alpha \beta} = \pa_{\alpha}\br\cdot \pa_{\beta}\br$ from the Euclidean metric $\delta_{\alpha \beta}$. The minimum of the energy occurs for a flat configuration which can be chosen as $\br = \bx$. The position vector $\br$ is conveniently parametrized by separating in-plane and out-of-plane fluctuation modes: $\br = (\bx + \bu, \bh)$, where $\bu \in \mathbb{R}^{D}$ and $\bh \in \mathbb{R}^{d-D}$~\cite{aronovitz_prl_1988}. An analysis of the canonical dimension of interactions shows that the model has $D = 4$ as the upper critical dimension~\cite{aronovitz_prl_1988, guitter_jpf_1989}. By dropping all interactions which are irrelevant by power counting near $D = 4$~\cite{aronovitz_prl_1988, guitter_jpf_1989, guitter_prl_1988}, the Hamiltonian reduces to the energy functional:
\begin{equation} \label{def:Hbar}
\bar{H} = \frac{1}{2}\int {\rm d}^{D}x \,\big[\kappa (\pa^{2}\bh)^{2} + \lambda u_{\alpha \alpha}^{2} + 2 \mu u_{\alpha \beta} u_{\alpha \beta}\big]~,
\end{equation}
with:
\begin{equation} \label{def:u}
u_{\alpha \beta} = \frac{1}{2} (\pa_{\alpha} u_{\beta} + \pa_{\beta} u_{\alpha} + \pa_{\alpha} \bh \cdot \pa_{\beta} \bh)~.
\end{equation}
The model defined by Eqs.~\eqref{def:Hbar} and~\eqref{def:u} has been investigated extensively~\cite{nelson_jpf_1987, guitter_jpf_1989, aronovitz_prl_1988, guitter_prl_1988, gornyi_prb_2015, bowick_pr_2001, burmistrov_aop_2018, le-doussal_prl_1992, le-doussal_aop_2018, gazit_pre_2009}. We will assume it as the starting point of the analysis in this work\footnote{The Hamiltonian $\bar{H}$, Eq.~\eqref{def:Hbar}, finds its rigorous justification as the most general local Hamiltonian allowed by power counting near the upper critical dimension $D =4$~\cite{aronovitz_prl_1988, guitter_jpf_1989}. In analogy with field-theoretic approaches to critical phenomena~\cite{zinn-justin_qft}, we can interpret Eq.~\eqref{def:Hbar} as an effective Hamiltonian suitable for calculations of the exponents of leading scaling behavior to all orders in the $\ve$-expansion. In this work, we assume that Eq.~\eqref{def:Hbar} holds directly in $D = 2$.}.
 
The Hamiltonian $\bar{H}$ is quadratic in the in-plane displacement vector $\bu$. The $\bu$ field can, therefore, be integrated out exactly~\cite{nelson_jpf_1987, nelson_statistical, aronovitz_jpf_1989, le-doussal_prl_1992, gornyi_prb_2015}. The resulting effective Hamiltonian for transverse fluctuations reads, in the physical case $D = 2$~\cite{nelson_jpf_1987, nelson_statistical}: 
\begin{equation} \label{eq:Heff}
H_{\rm eff} = \frac{1}{2} \int' \frac{{\rm d}^{2}q}{(2 \pi)^{2}} \Big[\kappa q^{4} |\bh(\bq)|^{2} + Y \lt|\frac{K(\bq)}{q^{2}}\rt|^{2} \Big]~.
\end{equation}
where $\bh(\bq)$ denotes Fourier components of $\bh(\bx)$, $K(\bq)$ the Fourier transform of the local operator
\begin{equation} \label{def:K}
K(\bx) = -\frac{1}{2}(\delta_{\alpha \beta} \pa^{2} - \pa_{\alpha} \pa_{\beta})(\pa_{\alpha} \bh \cdot \pa_{\beta} \bh) = \frac{1}{2}\big[(\pa^{2} \bh \cdot \pa^{2}\bh) - (\pa_{\alpha} \pa_{\beta} \bh \cdot \pa_{\alpha} \pa_{\beta} \bh)\big]~.
\end{equation}
and $Y = 4 \mu (\lambda + \mu)/(\lambda + 2 \mu)$ is the two dimensional Young modulus~\cite{katsnelson_graphene}. The composite field $K(\bx)$ is, at leading order for small deformations, the Gaussian curvature of the membrane~\cite{nelson_jpf_1987}. In the following, the effective theory defined by Eqs.~\eqref{eq:Heff} and~\eqref{def:K} will be referred to as 'Gaussian curvature interaction' model or for brevity GCI model.

A remark is in order about Eq.~\eqref{eq:Heff}. In the integration over in-plane displacement vectors $\bu$, zero modes of the strain tensor require a separate analysis (see Chap.~6 of Ref.~\cite{nelson_statistical}). Functional integration involves, besides finite-wavelength phonon deformations, a sum over macroscopic in-plane deformations of the crystal. As it was shown in Ref.~\cite{nelson_statistical}, this leads to a prescription on the momentum integral defining the effective Hamiltonian: the $\bq = 0$ contribution must be omitted from integration in Eq.~\eqref{eq:Heff} (see also Refs.~\cite{aronovitz_jpf_1989, le-doussal_aop_2018}).

While the GCI model is equivalent to Eq.~\eqref{def:Hbar} in two dimensions, integrating out in-plane phonons in a membrane of generic internal dimensionality leads to a different effective theory for transverse fluctuations. The analogue of Eq.~\eqref{eq:Heff} in  dimension $D \neq 2$ involves two independent non-local couplings, one of which characterized by tensor symmetry in internal space~\cite{aronovitz_jpf_1989, le-doussal_prl_1992, gornyi_prb_2015, le-doussal_aop_2018}. Explicitly, the effective Hamiltonian reads\footnote{In the notation of Ref.~\cite{le-doussal_aop_2018}, the interaction term in Eq.~\eqref{eq:HeffD} is rescaled by a factor $1/d_{c}$.}~\cite{le-doussal_prl_1992, le-doussal_aop_2018}:
\begin{equation} \label{eq:HeffD}
H^{(D)}_{\rm eff}= \frac{\kappa}{2}\int_{\bq} q^{4} \lt|\bh(\bq)\rt|^{2} + \frac{1}{4}\int_{\bk_{1}} \int_{\bk_{2}}\int_{\bk_{3}} R_{\alpha \beta, \gamma \delta}(\bq) k_{1 \alpha} k_{2 \beta} k_{3\gamma} k_{4 \delta}\lt(\bh(\bk_{1}) \cdot \bh(\bk_{2})\rt) \lt(\bh(\bk_{3}) \cdot \bh(\bk_{4})\rt)~,
\end{equation}
where $\bq = \bk_{1} + \bk_{2}$ and $ \bk_{1} + \bk_{2} + \bk_{3} + \bk_{4} = 0$, $\int_{\bk} = \int {\rm d}^{D}k/(2\pi)^{D}$ denotes momentum integration and
\begin{equation} \label{def:R}
R_{\alpha \beta, \gamma \delta}(\bq) =  \frac{\mu(D \lambda + 2\mu)}{(\lambda + 2\mu)} N_{\alpha \beta, \gamma \delta} + \mu M_{\alpha \beta, \gamma \delta}
\end{equation}
is defined in terms of the transverse projectors
\begin{equation} \label{def:M-N}
N_{\alpha \beta, \gamma \delta} = \frac{1}{D-1}P^{T}_{\alpha \beta} P^{T}_{\gamma \delta} ~,\qquad M_{\alpha \beta, \gamma \delta} = \frac{1}{2} \lt(P^{T}_{\alpha \gamma} P^{T}_{\beta \delta} + P^{T}_{\beta \delta} P^{T}_{\alpha \gamma}\rt) - N_{\alpha \beta, \gamma \delta}~,\qquad P^{T}_{\alpha \beta} = \delta_{\alpha \beta} - \frac{q_{\alpha} q_{\beta}}{q^{2}}~.
\end{equation}
For $D = 2$, the tensor $M_{\alpha \beta, \gamma \delta}$ vanishes identically and Eq.~\eqref{eq:HeffD} reduces to Eq.~\eqref{eq:Heff}. In this work we consider the GCI model, Eq.~\eqref{eq:Heff}, as a starting point for dimensional continuation. In the notation of Eq.~\eqref{eq:HeffD}, this is equivalent to setting $M_{\alpha \beta, \gamma \delta} = 0$, for all $D$, in the effective interaction $R_{\alpha \beta, \gamma \delta}(\bq)$, Eq.~\eqref{def:R}.

To analyze the GCI model, it is convenient to decouple interactions by a Hubbard-Stratonovich transformation:
\begin{equation}
{\rm e}^{-\frac{H_{\rm eff}[\bh(\bx)]}{T}} = \int [\mathscr{D}\lambda(\bx)] \,{\rm e}^{-\Ha[\bh(\bx), \lambda(\bx)]}~,
\end{equation}
where $\int [\mathscr{D}\lambda(\bx)] $ denotes functional integration over an auxiliary real field $\lambda(\bx)$ coupled to curvature $K(\bx)$. After rescaling field amplitudes, the Hamiltonian reads:
\begin{equation} \label{eq:HS}
\Ha =\int {\rm d}^{2}x\,\Big[\frac{1}{2}(\pa^{2}\bh)^{2} + \frac{1}{2 Y_{0}} (\pa^{2} \lambda)^{2} + i \lambda K \Big]~,
\end{equation}
where $Y_{0} = T Y/\kappa^{2}$.

\begin{figure}[ht]
\centering
\includegraphics[scale=1]{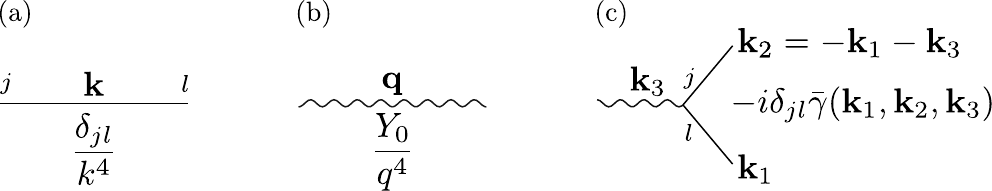}
\caption{\label{fig:feynman-rules} Bare propagators and vertex of the model. (a) $\bh$-field propagator, (b) $\lambda$-field propagator, (c) interaction vertex.}
\end{figure}

Equilibrium correlation functions at temperature $T$ are calculated by functional integration over $\bh(\bx)$ and $\lambda(\bx)$ with the statistical weigth $\exp(-\Ha)$~\cite{zinn-justin_qft}. The corresponding perturbation theory is generated by the propagators and the vertex illustrated in Fig.~\ref{fig:feynman-rules}. The vertex presents a simple geometric structure. From the definition of the curvature $K(\bx)$, Eq.~\eqref{def:K} and momentum conservation, $\bk_{1} + \bk_{2} + \bk_{3}=0$, it can be verified that:
\begin{equation} \label{def:gamma}
\bar{\gamma}(\bk_{1}, \bk_{2}, \bk_{3}) = k_{1}^{2} k_{2}^{2} - (\bk_{1} \cdot \bk_{2})^{2} = k_{2}^{2} k_{3}^{2} - (\bk_{2}\cdot \bk_{3})^{2} = k_{3}^{2} k_{1}^{2} - (\bk_{3} \cdot \bk_{1})^{2}~.
\end{equation}

\section{Renormalization group equations} \label{sec:RG}
Representation of the GCI model by Hubbard-Stratonovich transformation gives access to the methods of perturbative renormalization of local field theories~\cite{zinn-justin_qft, amit_qft, parisi_sft}. This section presents a derivation of renormalization group equations for the theory in the representation expressed by Eqs.~\eqref{eq:HS} and~\eqref{def:K}.

By dimensional continuation of Feynman diagrams~\cite{zinn-justin_qft}, the GCI model can be extended to arbitrary non-integer dimension $D$. Power counting shows that the coupling constant $Y_{0}$ has dimension $\ve = (4-D)$.  The GCI theory has therefore the same upper critical dimension $D_{\rm uc} = 4$ as the conventional model of $D$-dimensional elastic membranes~\cite{aronovitz_prl_1988}. For $D < 4$ the perturbative expansion breaks down at sufficiently large order because of infrared divergences, similarly to the theory of critical behavior~\cite{zinn-justin_qft, parisi_sft}. In analogy with earlier approaches to crystalline membranes~\cite{guitter_jpf_1989, aronovitz_prl_1988, guitter_prl_1988} and critical phenomena, the GCI model can be analyzed within an $\ve = (4-D)$-expansion~\cite{zinn-justin_qft}, in which perturbative corrections are logarithmic and massless perturbation theory is well-defined to all orders. 

\begin{figure}[ht]
\centering
\includegraphics[scale=1]{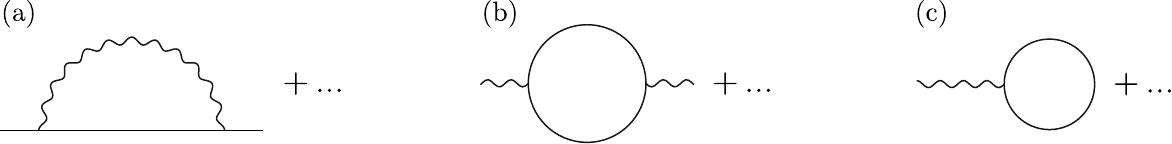}
\caption{\label{fig:superficial-divergences} Superficially divergent 1PI correlation functions.}
\end{figure}

A power counting analysis of one-particle irreducible (1PI) correlation functions shows that the $\bh$ and $\lambda$ field self-energies and the $\lambda$ one-point function are the only superficial ultraviolet divergences~\cite{zinn-justin_qft} in four dimensions. The structure of the vertex, Eq.~\eqref{def:gamma}, implies that in any diagram two powers of momentum can be attached to each external leg and factorized from the loop momentum integration\footnote{This property reflects the symmetry of the theory under the shifts $\bh(\bx)\to \bh(\bx) + \bm{A}_{\alpha} x_{\alpha} + \bm{B}$ and $\lambda(\bx) \to \lambda(\bx) + A'_{\alpha}x_{\alpha} + B'$, which leave the Hamiltonian~\eqref{eq:HS} invariant up to boundary terms. Shifts of $\bh(\bx)$ by a first order polynomial in the coordinates correspond to translations and rotations of the membrane in the $d$-dimensional embedding space. Invariance of the theory is therefore related to the Goldstone-mode character of the out-of-plane fluctuation fields and to Ward identities of embedding space rotational invariance. The analysis is consistent with Refs.~\cite{le-doussal_prl_1992, le-doussal_aop_2018, aronovitz_jpf_1989}.}. For any 1PI diagram $\Delta$ with $L$ loops, $V$ vertices, $I$ internal and $E$ external lines (of either solid or wiggly type), the degree of superficial divergence~\cite{zinn-justin_qft} is therefore
\begin{equation}
\delta_{\Delta} = D L + 4 V - 4 I - 2 E~, 
\end{equation}
or equivalently, using the topological relations $L = 2 V - I - E + 1$ and  $2I + E = 3V$,
\begin{equation}
\delta_{\Delta} = D - \frac{D}{2} E - \frac{1}{2}(4-D) V ~.
\end{equation}
For $D = 4$ the degree of divergence is independent on the number of vertices and the model is renormalizable by power counting. The only superficially divergent 1PI functions, characterized by $\delta_{\Delta} > 0 $, are two-point functions in Fig.~\ref{fig:superficial-divergences}(a) and~\ref{fig:superficial-divergences}(b), which diverge logarithmically, and~\ref{fig:superficial-divergences}(c), which diverges quadratically. The three-point vertex function is, instead, convergent and it does not require the introduction of an independent renormalization constant\footnote{\label{fn:vertex-convergence} The convergence of the vertex function and the subsequent absence of vertex renormalization was appreciated  in Refs.~\cite{le-doussal_prl_1992, le-doussal_aop_2018}. It was related to Ward identities and it was proven to imply the exactness of the self-consistent screening approximation to O($\ve$) in the $\ve$-expansion for $D$-dimensional elastic membranes~\cite{le-doussal_prl_1992, le-doussal_aop_2018}.}.

The one-point function (Fig.~\ref{fig:superficial-divergences}c) vanishes at nonzero momentum. On the other hand, the field $\lambda$ was introduced to mediate the non-local interaction of Eq.~\eqref{eq:Heff}, in which the $\bq = 0$ mode is excluded. We thus assume that functional integration runs only over finite-momentum Fourier components of the $\lambda$ field. Diagrams for the one-point function can then be dropped, consistently with the elimination of tadpole diagrams in earlier approaches to crystalline membranes~\cite{aronovitz_jpf_1989, le-doussal_aop_2018} (see also Chap.~6 of Ref.~\cite{nelson_statistical}).

Ultraviolet divergences are therefore removed by the introduction of two renormalization constants, corresponding to the superficial divergences in the two-point functions~\cite{le-doussal_prl_1992, le-doussal_aop_2018}. Within dimensional regularization and the minimal subtraction scheme~\cite{zinn-justin_qft, amit_qft}, the renormalized Hamiltonian equipped with necessary counterterms reads:
\begin{equation} \label{eq:Hren}
\tilde{\Ha} = \int {\rm d}^{D}x \,\bigg[\frac{Z}{2} (\pa^{2} \bh)^{2} + \frac{s_{D}}{2 M^{\ve} Z_{y} y} (\pa^{2} \lambda)^{2} + i \lambda K\bigg]~,
\end{equation}
where $M$ is an arbitrary wavevector scale and $y$ is a dimensionless renormalized coupling. For later convenience in explicit calculations, we have redefined the coupling constant by introducing $y_{0} = s_{D} Y_{0}$, with
\begin{equation}
s_{D} = \frac{(D^{2}-1)\Gamma^{2}(D/2)\Gamma(3 - D/2)}{4 (4\pi)^{D/2} \Gamma(D)}~,
\end{equation}
because a geometric factor similar to $s_{D}$ is generated in the one-loop Feynman  diagrams~\cite{le-doussal_prl_1992, le-doussal_aop_2018}. In the renormalization of the theory, this parametrization converts minimal subtraction into an analogue of the modified minimal subtraction scheme~\cite{zinn-justin_qft}. The amplitudes $Z$ and $Z_{y}$ are double series in $y$ and $1/\ve$ and, being dimensionless, do not depend explicitly on $M$. Comparison of Eq.~\eqref{eq:HS} and Eq.~\eqref{eq:Hren} gives the following relations between bare and renormalized quantities:
\begin{equation}
\bh(\bx) = \sqrt{Z} \,\tilde{\bh}(\bx)~,\qquad \lambda(\bx) = \frac{1}{Z} \tilde{\lambda}(\bx)~, \qquad y_{0} = \frac{M^{\ve} Z_{y}}{Z^{2}}y~,\qquad \Ha[\bh(\bx), \lambda(\bx)] = \tilde{\Ha}[\tilde{\bh}(\bx), \tilde{\lambda}(\bx)]~.
\end{equation}
The renormalization of 1PI correlation functions with $n$ external $\bh$ lines and $\ell$ external $\lambda$ lines reads:
\begin{equation}
\Gamma_{i_{1} .. i_{n}}^{(n, \ell)}(\bk_{1}, .., \bk_{n}; \bq_{1}, .., \bq_{\ell}; y_{0}) = Z^{\ell - \frac{n}{2}} \tilde{\Gamma}_{i_{1} .. i_{n}}^{(n, \ell)} (\bk_{1}, .., \bk_{n}; \bq_{1}, .., \bq_{\ell}; M, y)~.
\end{equation}
Renormalization group equations follow, in a standard way~\cite{zinn-justin_qft, amit_qft}, from the independence of the bare functions $\Gamma^{(n, \ell)}$ on the wavevector scale $M$. Introducing
\begin{equation} \label{def:RG-functions}
\beta(y) = \frac{\pa y}{\pa \ln M}\bigg|_{y_{0}}~,\qquad \bar{\eta}(y) = \frac{\pa \ln Z}{\pa \ln M}\bigg|_{y_{0}}~,
\end{equation}
the RG equations for one-particle irreducible correlation functions read:
\begin{equation}
\Big[M\frac{\pa}{\pa M}  + \beta(y) \frac{\pa}{\pa y} - \lt(\frac{n}{2} - \ell\rt)\bar{\eta}(y)\Big] \tilde{\Gamma}_{i_{1} .. i_{n}}^{(n, \ell)}(\bk_{1}, .., \bk_{n}; \bq_{1}, .., \bq_{\ell}; M, y) = 0~.
\end{equation}
As a consequence of dimensional regularization and the minimal subtraction prescription~\cite{zinn-justin_qft}, $\bar{\eta}(y)$ does not depend explicitly on $\ve$ and $\beta(y) = -\ve y + b(y)$ where $b(y)$ is $\ve$-independent. Being dimensionless, $\beta(y)$ and $\bar{\eta}(y)$ are independent of the renormalization scale $M$.

In the long wavelength limit, the running coupling approaches an infrared-attractive fixed point $y = y_{*}$, determined by the condition $\beta(y_{*})=0$~\cite{zinn-justin_qft, amit_qft}. At the fixed point, a combination of RG equations and dimensional analysis shows that correlation functions\footnote{In Eq.~\eqref{eq:scaling-relation}, 1PI correlation functions in momentum space are defined after factorization and cancellation of the momentum-conservation factor $(2\pi)^{D} \delta\big(\sum_{i} \bk_{i} + \sum_{j} \bq_{j}\big)$.} scale according to:
\begin{equation} \label{eq:scaling-relation}
\tilde{\Gamma}_{i_{1} .. i_{n}}^{(n, \ell)}(\rho\bk_{1}, .., \rho\bk_{n}; \rho\bq_{1}, .., \rho\bq_{\ell}; M, y_{*}) = \rho^{D + \frac{n}{2}(4 - D - \eta) + \eta \ell} \tilde{\Gamma}_{i_{1} .. i_{n}}^{(n, \ell)}(\bk_{1}, .., \bk_{n}; \bq_{1}, .., \bq_{\ell}; M, y_{*})~,
\end{equation}
where the scaling exponent is $\eta = \bar{\eta}(y_{*})$. In particular two-point functions satisfy the scaling relations:
\begin{equation} \label{eq:scaling-two-point-functions}
 \tilde{\Gamma}_{ij}^{(2, 0)}(\bk)  \propto \delta_{ij}|\bk|^{4 - \eta}~,\qquad \tilde{\Gamma}^{(0, 2)}(\bq) \propto \lt|\bq\rt|^{D + 2 \eta}~.
\end{equation}
We can interpret Eq.~\eqref{eq:scaling-two-point-functions} by recognizing a power law divergence of the effective bending rigidity $\kappa_{\rm R}(q)$ and a suppression of the effective Young modulus $Y_{\rm R}(q)$. Defining $\kappa_{\rm R}(q)$ and $Y_{\rm R}(q)$ by the identifications 
\begin{equation} \label{eq:renormalized-elastic-coefficients}
\Gamma^{(2, 0)}_{ij}(\bq) = \delta_{ij}\frac{\kappa_{\rm R}(q) q^{4}}{\kappa} ~, \qquad \Gamma^{(0, 2)}(\bq) = \frac{\kappa^{2} q^{4}}{T Y_{\rm R}(q)}~,
\end{equation}
implies, since bare and renormalized correlation functions are proportional, that $\kappa_{\rm R}(q) \propto q^{-\eta}$ and $Y_{\rm R}(q) \propto q^{\eta_{u}}$ with $\eta_{u} = 4 - D - 2 \eta$. The relation between $\eta_{u}$ and $\eta$ is consistent with the theory of $D$-dimensional membranes, for which a well-known analogue exponent identity holds in arbitrary dimension~\cite{aronovitz_jpf_1989, aronovitz_prl_1988, guitter_prl_1988, le-doussal_prl_1992, le-doussal_aop_2018, gazit_pre_2009}.

\section{Scaling exponent: second-order $\ve$-expansion}
This section reports explicit results for the scaling exponent $\eta$ to order $\ve^{2}$ in the $\ve$-expansion. As it will be verified, the coupling strength $y_{*}$ at the fixed point is of order $\ve$ near dimension four. Determination of $\eta$ with accuracy $\ve^{2}$, therefore, requires the knowledge of $\beta(y)$ and $\bar{\eta}(y)$ to order $y^{3}$ and $y^{2}$ respectively. RG functions at this order can be calculated by computing the renormalization constants $Z$ and $Z_{y}$ at two-loop level.

\begin{figure}[ht]
\centering
\includegraphics[scale=1]{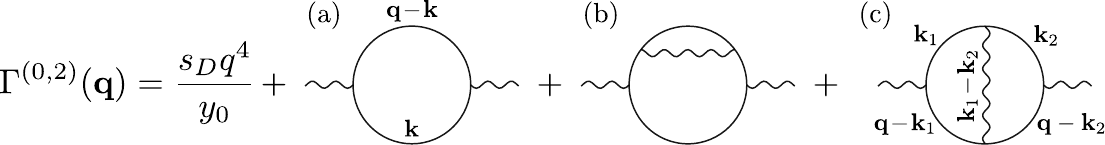} 

\vspace{0.8cm}

\includegraphics[scale=1]{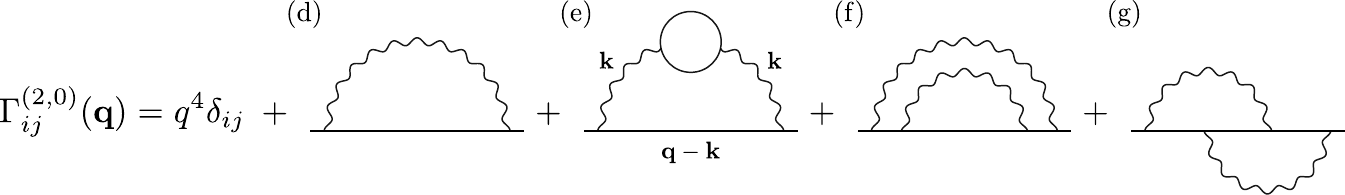}
\caption{\label{fig:two-loop-feynman} Feynman diagrams for the one-particle irreducible two-point functions at order two-loops. The values of diagrams (a), (b) and (c) are denoted as $D_{a}(\bq)$, $D_{\rm b}(\bq)$ and  $D_{\rm c}(\bq)$ respectively. Diagrams (d), (e), (f) and (g) are proportional to the identity in $d_{c}$-dimensional space and, therefore, are denoted as $\delta_{ij} D_{l}(\bq)$, where $l$ = d, .., g.} 
\end{figure}

The bare one-particle irreducible two-point functions are given, in second order perturbation theory, by the sum of Feynman diagrams in Fig.~\ref{fig:two-loop-feynman}.

Perturbative calculations are extensively simplified by the similarity between $\bh$- and $\lambda$-field propagators, both scaling with momentum $\bk$ as $k^{-4}$, and the permutation-invariance of the vertex function expressed by Eq.~\eqref{def:gamma}. These properties imply that, up to a  prefactor, the loop integral corresponding to a Feynman diagram does not depend on the type (solid or wavy) of its lines, but only on its overall connectivity. Denoting as $D_{l}(\bq)$ ($l$ = a, .., c) and $\delta_{ij} D_{l}(\bq)$ ($l$=d, .. , g) the values of the $l$-th Feynman diagram in Fig.~\ref{fig:two-loop-feynman}, the following relations hold\footnote{In Eqs.~\eqref{eq:relation-diagrams} and~\eqref{eq:diagrams-integrals}, combinatorial factors $1/2$ appear in presence of closed solid-line loops with flip symmetry. An analogue Feynman rule applies to the the Ginzburg-Landau model in a Hubbard-Stratonovich representation~\cite{ma_qft}.}:
\begin{equation} \label{eq:relation-diagrams}
\frac{y_{0}}{s_{D}} D_{\rm a}(\bq) = \frac{d_{c}}{2}D_{\rm d}(\bq)~,\qquad D_{\rm e}(\bq)  = \frac{y_{0}}{2 s_{D}} D_{\rm b}(\bq) = \frac{d_{c}}{2} D_{\rm f}(\bq)~,\qquad  \frac{y_{0}}{s_{D}} D_{\rm c}(\bq) = \frac{d_{c}}{2} D_{\rm g}(\bq)~.
\end{equation}
Calculation of the seven diagrams thus reduces to the computation of three independent integrals. Explicit expressions for diagrams (a), (e) and (c) are:
\begin{equation} \label{eq:diagrams-integrals}
\begin{gathered}
D_{\rm a}(\bq) = \frac{d_{c}}{2} \int \frac{{\rm d}^{D}k}{(2\pi)^{D}} \frac{(q^{2} k^{2} - (\bq \cdot \bk)^{2})^{2}}{|\bk|^{4} |\bq - \bk|^{4}}~,\\
D_{\rm e}(\bq) = - \int \frac{{\rm d}^{D}k}{(2\pi)^{D}} \frac{(q^{2} k^{2} - (\bq \cdot \bk)^{2})^{2}}{|\bq - \bk|^{4}} \bigg[\frac{y_{0}}{s_{D} k^{4}} D_{\rm a}(\bk)\frac{y_{0}}{s_{D} k^{4}}\bigg]~,\\
D_{\rm c}(\bq) = -\frac{d_{c} y_{0}}{2 s_{D}} \int \frac{{\rm d}^{D}k_{1}}{(2\pi)^{D}} \int \frac{{\rm d}^{D}k_{2}}{(2\pi)^{D}} \frac{\gamma(\bq, \bk_{1}) \gamma(\bk_{1}, \bk_{2}) \gamma(\bq, \bk_{2}) \gamma(\bq-\bk_{1}, \bq-\bk_{2})}{|\bk_{1}|^{4} |\bk_{2}|^{4} |\bq - \bk_{2}|^{4}|\bq - \bk_{1}|^{4} |\bk_{1} - \bk_{2}|^{4}}~,
\end{gathered}
\end{equation}
where $\gamma(\bk_{1}, \bk_{2}) = k_{1}^{2} k_{2}^{2} - (\bk_{1} \cdot \bk_{2})^{2}$. The diagrams $D_{\rm a}$ and $D_{\rm e}$ can be deduced from the general integral~\cite{le-doussal_aop_2018}
\begin{equation} \label{eq:scsa-integral}
\Pi(\eta, \eta', D)  = \int \frac{{\rm d}^{D}k}{(2\pi)^{D}} \frac{(k^{2} - (\hq \cdot \bk)^{2})^{2}}{|\bk|^{4 - \eta} |\bk + \hq|^{4-\eta'}} = (D^{2}-1) \frac{\Gamma(2 - \frac{\eta + \eta'}{2}- \frac{D}{2}) \Gamma(\frac{D}{2} + \frac{\eta}{2})\Gamma(\frac{D}{2} + \frac{\eta'}{2})}{4 (4\pi)^{D/2} \Gamma(2 - \frac{\eta}{2}) \Gamma(2 - \frac{\eta'}{2})\Gamma(D + \frac{\eta + \eta'}{2})}~,
\end{equation}
where $\hq=\bq/|\bq|$ is an unit vector.  Using Eq.~\eqref{eq:scsa-integral} repeatedly gives the expressions for the diagrams
\begin{equation}
\begin{gathered}
D_{\rm a}(\bq) = \frac{d_{c}}{2} \Pi(0, 0, D) q^{4-\ve} = \frac{d_{c}s_{D}}{\ve} q^{4-\ve}~,\\
D_{\rm e}(\bq) = - \frac{d_{c}}{2}  \Pi(0, 0, D) \Pi(-\ve, 0, D) \frac{y_{0}^{2} q^{4 - 2 \ve}}{s_{D}^{2}}~.
\end{gathered}
\end{equation}
Near $D = 4$, $D_{\rm e}(\bq)$ has the expansion:
\begin{equation}
D_{\rm e}(\bq) = -d_{c}\lt(\frac{1}{\ve^{2}} - \frac{1}{12 \ve} + {\rm O}(1)\rt) y_{0}^{2} q^{4 - 2 \ve}~,
\end{equation}
where ${\rm O}(1)$ denotes the finite part of $D_{\rm e}(\bq)$ for $\ve \to 0$. The remaining independent diagram $D_{\rm c}(\bq)$ gives, by dimensional analysis:
\begin{equation}
D_{\rm c}(\bq) = - \frac{d_{c} s_{D} a_{D}}{2}  y_{0} q^{4 - 2\ve}~,
\end{equation}
where $a_{D}$ is a function of internal dimension $D$. As it is shown in~\ref{app:two-loop-self-energy}, $a_{D}$ presents a first-order pole at $D = 4$
\begin{equation} \label{eq:aD-pole}
a_{D} =  \frac{A}{\ve} + {\rm O}\lt(1\rt)~,
\end{equation}
with residue  $A = 121/90$, corresponding to the UV divergence of $D_{\rm c}(\bq)$ in four dimensions. The bare two-point functions can therefore be written as:
\begin{equation}
\begin{split}
\Gamma^{(0, 2)}(\bq)  & = \frac{s_{D} q^{4}}{y_{0}}   +   D_{\rm a}(\bq) + D_{\rm b}(\bq) + D_{\rm c}(\bq)  \\ & = \frac{s_{D} q^{4}}{y_{0}} \bigg[1 + \frac{d_{c}}{\ve}\frac{y_{0}}{q^{\ve}} - d_{c} \lt(\frac{2}{\ve^{2}} - \frac{1}{6 \ve} + \frac{A}{2\ve}+ {\rm O}(1)\rt) \frac{y_{0}^{2}}{q^{2\ve}} +  {\rm O}\lt(y_{0}^{3}\rt)\bigg]~,
\end{split}
\end{equation}
\begin{equation}
\begin{split}
\Gamma^{(2, 0)}(\bq) & = \delta_{ij} [q^{4}  + D_{\rm d}(\bq) + D_{\rm e}(\bq) + D_{\rm f}(\bq) + D_{\rm g}(\bq)] \\
& = \delta_{ij} q^{4} \bigg[1 + \frac{2}{\ve} \frac{y_{0}}{q^{\ve}} - \bigg(1 + \frac{d_{c}}{2}\bigg)\bigg(\frac{2}{\ve^{2}} -\frac{1}{6 \ve} + {\rm O}(1)\bigg) \frac{y_{0}^{2}}{q^{2\ve}} - \lt(\frac{A}{\ve} + {\rm O}(1)\rt) \frac{y_{0}^{2}}{q^{2\ve}} + {\rm O}\lt(y_{0}^{3}\rt)\bigg]~.
\end{split}
\end{equation}
With the choice of renormalization constants
\begin{equation}
\begin{gathered}
 Z_{y} = 1 + \frac{d_{c}y}{\ve} +  d_{c} (d_{c} + 2)\frac{ y^{2}}{\ve^{2}} + \frac{d_{c}}{2} \lt(\frac{1}{3} - A\rt) \frac{y^{2}}{\ve} + {\rm O}\lt(y^{3}\rt)~,\\
 Z = 1 - \frac{2 y}{\ve} - (d_{c} + 2) \frac{y^{2}}{\ve^{2}} - \lt(\frac{d_{c} + 2}{12} - A\rt)\frac{y^{2}}{\ve} + {\rm O}\lt(y^{3}\rt) ~,
\end{gathered}
\end{equation}
the renormalized correlation functions $\tilde{\Gamma}^{(0, 2)} = Z^{-2} \Gamma^{(0, 2)}$ and $\tilde{\Gamma}^{(2, 0)} = Z \Gamma^{(2, 0)}$ are finite to order $y^{2}$ after the coupling renormalization $y_{0} = M^{\ve} Z_{y}y/Z^{2} $. The corresponding RG functions can be determined from the relations~\cite{zinn-justin_qft}
\begin{equation}
 \beta(y) = \frac{-\ve y}{1 + \frac{\pa \ln (Z_{y}/Z^{2})}{\pa \ln y}}~,\qquad \bar{\eta}(y) = \beta(y) \frac{\pa \ln Z}{\pa y}~,
\end{equation}
which lead to:
\begin{gather}
\beta(y) = -\ve y + (d_{c} + 4) y^{2}  +  \bigg(\frac{2}{3} (d_{c} + 1) - (d_{c} + 4)A \bigg) y^{3} + {\rm O}\lt(y^{4}\rt)~, \label{eq:beta}\\
\bar{\eta}(y) = 2 y + \lt(\frac{d_{c} + 2}{6} - 2A\rt) y^{2} + {\rm O}\lt(y^{3}\rt)\label{eq:eta}~.
\end{gather}
The $\beta$ function describes, for $\ve$ small, a renormalization group flow from the Gaussian fixed point $y = 0$ to a nontrivial infrared stable fixed point $y = y_{*}$ corresponding to the coupling strength
\begin{equation}
y_{*} = \frac{\ve}{d_{c} + 4} - \frac{\lt(\frac{2}{3}(d_{c} + 1) - (d_{c} + 4)A\rt)}{\lt(d_{c} + 4\rt)^{3}}\ve^{2} + {\rm O}(\ve^{3})~.
\end{equation}
The anomalous dimension controlling the long-wavelength scaling behavior is therefore
\begin{equation} \label{eq:scaling-exponent}
\eta = \bar{\eta}(y_{*}) = \frac{2\ve}{d_{c} + 4} -  \frac{d_{c}(2 - d_{c})}{6 (d_{c} + 4)^{3}} \ve^{2} + {\rm O}(\ve^{3})~.
\end{equation}

\subsection{Exactness of the self-consistent screening approximation to O$(\ve^{2})$.} \label{sec:SCSA}

As Eq.~\eqref{eq:scaling-exponent} shows, the scaling exponent $\eta$ is insensitive to the value of the residue $A$ describing the contribution of diagrams (c) and (g) in Fig.~\ref{fig:two-loop-feynman}. Only diagrams (a), (b), (d), (e) and (f), representing propagator corrections contribute to the value of $\eta$ to order $\ve^{2}$. This suggests that the self-consistent screening approximation~\cite{le-doussal_prl_1992, le-doussal_aop_2018} is exact to O($\ve^{2}$) for the GCI model.

\begin{figure}[ht]
\centering
\includegraphics[scale=1]{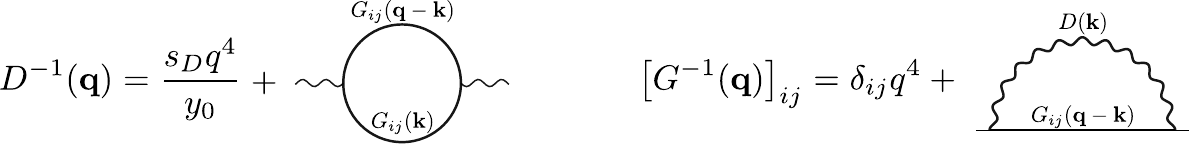}
\caption{\label{fig:scsa} Diagrams corresponding to SCSA equations. $D(\bq)$ and $G_{ij}(\bq)$ denote the $\lambda$- and $\bh$-field propagators respectively.}
\end{figure}

Within the SCSA, scaling exponents are determined by identifying power-law solutions to truncated Dyson equations for the $\lambda$- and $\bh$- field propagators $D(\bq)$ and $G_{ij}(\bq)$. In analogy with the theory of $D$-dimensional membranes~\cite{le-doussal_prl_1992, le-doussal_aop_2018}, we define the self-consistent screening approximation by the equations
\begin{equation} \label{eq:scsa}
\begin{gathered}
\lt[G^{-1}(\bq)\rt]_{ij} = \delta_{ij} q^{4} + \int \frac{{\rm d}^{D}k}{(2\pi)^{D}} \lt(q^{2} k^{2} - (\bq \cdot \bk)^{2}\rt)^{2} G_{ij}(\bk) D(\bq - \bk)~,\\
D^{-1}(\bq) = \frac{s_{D} q^{4}}{y_{0}} + \frac{1}{2}\int \frac{{\rm d}^{D}k}{(2\pi)^{D}}\lt(q^{2} k^{2} - (\bq\cdot \bk)^{2}\rt)^{2} G_{ij}(\bk) G_{ji}(\bq - \bk)~,
\end{gathered}
\end{equation}
which correspond to the diagrams in Fig.~\ref{fig:scsa}.

The inverse two-point functions $[G^{-1}_{ij}(\bq)]$ and $D^{-1}(\bq)$ approximate, within the SCSA,  the interacting 1PI two-point functions $\Gamma^{(2, 0)}_{ij}(\bq)$ and  $\Gamma^{(0, 2)}(\bq)$ respectively. In the long-wavelength, strong-coupling limit zero-order propagators are negligible compared to self-energy terms~\cite{gazit_pre_2009}. The SCSA equations admit, in this regime, scaling solutions of the form~\cite{le-doussal_prl_1992, le-doussal_aop_2018}:
\begin{equation}
G_{ij}(\bq) = z\delta_{ij} |\bq|^{-4 + \eta'}~,\qquad D(\bq) = c(\eta, D) z^{-2}|\bq|^{-4 + \eta'_{u}}~,
\end{equation}
where $z$ is a non-universal amplitude. Consistency of the solution with Eq.~\eqref{eq:scsa} imposes the exponent relation $\eta'_{u} = 4 - D - 2\eta'$ and an equation determining $\eta'$:
\begin{equation} \label{eq:scsa-igc}
\frac{d_{c}}{2} = \frac{\Pi(\eta', \eta'_{u}, D)}{\Pi(\eta', \eta', D)}=\frac{\Gamma(\frac{\eta'}{2}) \Gamma(2- \eta') \Gamma(D + \eta') \Gamma(2 - \frac{\eta'}{2})}{\Gamma(\frac{D}{2} + \frac{\eta'}{2}) \Gamma(2 - \eta'  - \frac{D}{2})\Gamma(\frac{D}{2} + \eta')   \Gamma(\frac{D}{2} + 2 - \frac{\eta'}{2})}~.
\end{equation}
The power-law behavior and the relation between $\eta_{u}'$ and $\eta'$ agrees with the scaling form of the effective bending rigidity and Young modulus, Eq.~\eqref{eq:renormalized-elastic-coefficients}. Solving Eq.~\eqref{eq:scsa-igc} order by order in $\ve$, it can be explicitly verified that the SCSA exponent $\eta'$ agrees with the exact $\ve$-expansion, Eq.~\eqref{eq:scaling-exponent}, not only at leading order but also at order $\ve^{2}$: $\eta' - \eta = {\rm O}(\ve^{3})$. The exactness of the SCSA at leading order in $\ve$ is a consequence of the structure of renormalization in the theory. Due to the absence of vertex renormalization, the one-loop RG functions are determined by diagrams without vertex corrections: the same diagrams included in the self-consistent screening approximation. An analogue situation occurs in the theory of $D$-dimensional crystalline membranes in the conventional dimensional continuation scheme for which the SCSA approximation yields the exact exponent $\eta = \ve/(2 + d_{c}/12)$~\cite{aronovitz_prl_1988} at leading order in the $\ve$-expansion~\cite{le-doussal_prl_1992, le-doussal_aop_2018}.

Exactness of the SCSA at next-to-leading order, which we have verified in the framework of the GCI model, follows instead from Eq.~\eqref{eq:relation-diagrams}, which relates the amplitude of diagrams for the $\bh$- and $\lambda$-field correlation functions. These relations can be traced to the permutation invariance of the vertex function, Eq.~\eqref{def:gamma}, and the identity (up to a factor $y_{0}/s_{D}$ and an identity matrix $\delta_{ij}$) of the $\bh$- and the $\lambda$-field non-interacting propagators.

As a final remark, we note that the SCSA equation for the GCI model, Eq.~\eqref{eq:scsa-igc}, is very similar to the self-consistent screening approximation for a crystalline $D$-dimensional membrane in the conventional dimensional continuation scheme~\cite{le-doussal_prl_1992, le-doussal_aop_2018}. The two equations differ by a simple factor $D(D-1)/2$, which reduces to unity in the physical case $D  = 2$. As expected, in two dimensions the GCI model and the theory of $D$-dimensional crystalline membranes present the same exponent $\eta' = 4/(d_{c} + \sqrt{16 - 2 d_{c} + d_{c}^{2}})$, approximately equal to $0.821$ for $d_{c} = 1$~\cite{le-doussal_prl_1992, le-doussal_aop_2018}.

\subsection{Extrapolation to the physical dimensionality}

For the physical codimension $d_{c} = 1$, the $\ve$-expansion of the scaling exponent, Eq.~\eqref{eq:scaling-exponent}, reduces to:
\begin{equation}
\eta = \frac{2 \ve}{5} - \frac{\ve^{2}}{750} + {\rm O}(\ve^{3})~.
\end{equation}
Compared to the leading order result, the O($\ve^{2}$) correction is strongly suppressed by its small numerical prefactor. Although $\ve$ is as large as $2$, a direct extrapolation to the physical internal dimension $D = 2$ reveals an unexpectedly small deviation between the first and the second order results, $\eta = 0.8$ and $\eta \simeq 0.795$ respectively.

The proximity of these results to previous values reported in the literature gives a further indication in support of the stability and accuracy of the low-order $\ve$-expansion. The exponent prediction is close to non-perturbative RG results $\eta \simeq 0.85$~\cite{kownacki_pre_2009, braghin_prb_2010, hasselmann_pre_2011}, the first and second-order SCSA $\eta \simeq 0.821$~\cite{le-doussal_prl_1992, le-doussal_aop_2018}, $\simeq 0.789$~\cite{gazit_pre_2009} and to several numerical simulations, reporting approximately $\eta \simeq 0.750$~\cite{bowick_pr_2001}, $ 0.81$~\cite{gompper_cis_1997, gompper_jpcm_1997}, $ 0.85$~\cite{los_prb_2009, wei_jcp_2014}, $ 0.795$~\cite{troster_prb_2013}, $ 0.793$--$0.795$~\cite{troster_pre_2015}, $ 0.66$~\cite{los_prb_2017}, $0.85$--$0.88$~\cite{hasik_prb_2018} (see Refs.~\cite{le-doussal_aop_2018, gompper_cis_1997, gompper_jpcm_1997} for results of early simulations). Finally, the leading-order extrapolation $\eta = 0.8$ is in exact agreement with the prediction of Ref.~\cite{kosmrlj_prb_2016}, where the exponent was obtained by a one-loop momentum shell renormalization group directly in $D = 2$.

To appreciate the closeness of the calculated exponent to earlier results, we note that the conventional $\ve$-expansion~\cite{aronovitz_prl_1988, aronovitz_jpf_1989, guitter_prl_1988, guitter_jpf_1989}, based on direct dimensional continuation of elasticity theory, gives $\eta = \ve/(2 + d_{c}/12)= 0.96$ after extrapolation to physical dimensionality. The deviation between the conventional and the GCI model $\ve$-expansions is amplified in the scaling exponent $\eta_{u} = 4 - D - 2\eta$. Extrapolation to $D = 2$ and $d_{c}=1$ gives $\eta_{u} = 0.08$ in the former scheme and $\eta_{u} = 0.4$, $\eta_{u} = 0.411$ in the first and second approximation of the latter. Comparison with the SCSA result $\eta_{u} \simeq 0.358$~\cite{le-doussal_prl_1992, le-doussal_aop_2018} indicates a closer agreement of GCI model results.

Despite the a posteriori agreement of second order results, the quantitative accuracy of the $\ve$-expansion at the physical dimension requires a more detailed investigation.

To conclude, we note a related analysis in Ref.~\cite{gazit_pre_2009}, which studied  extensions of the SCSA for physical 2D membranes. Among candidate extensions of the SCSA to higher orders, Ref.~\cite{gazit_pre_2009} considered a self-consistent two-loop theory, constructed by promoting propagator diagrams of one- and two-loop order to truncated Dyson equations. By a numerical analysis, Ref.~\cite{gazit_pre_2009} reported on the impossibility to determine self-consistent power-law solutions of the equations for $D =2$. Results obtained in this paper are complementary, being based on an expansion about $D = 4$. As it was shown in Sec.~\ref{sec:SCSA}, the leading-order SCSA remains valid to order $\ve^{2}$ near $D = 4$, without corrections from diagrams (c) and (g) in Fig.~\ref{fig:two-loop-feynman}. 

\section{Summary and conclusions}

In summary, this work examined a widely applied effective theory for out-of-plane fluctuations in 2D elastic membranes, controlled by bending rigidity and long-range interactions between Gaussian curvatures (here named GCI model). After dimensional continuation to non-integer internal dimension $D$, we analyzed the theory by systematic field-theory methods. An explicit calculation of RG functions at two-loop order was performed, leading to the quantitative prediction of the scaling exponent $\eta$ at the second, next-to-leading, order in an expansion in $\ve = 4 - D$. Generalization of the GCI model to non-integer dimension $D$ defines an alternative dimensional continuation of the theory of crystalline membranes, inequivalent to the direct continuation of $D$-dimensional elasticity theory, considered in previous $\ve$-expansion approaches.

The explicit result for the exponent, Eq.~\eqref{eq:scaling-exponent}, shows that $\eta$ is insensitive to vertex-correction diagrams up to O($\ve^{2}$) in the $\ve$-expansion. This property implies that, in the framework of the GCI model, the self-consistent screening approximation is exact to order $\ve^{2}$. While the exactness of the SCSA at first order is related to the structure of renormalization of the theory and, ultimately, to rotational symmetry in the embedding space, exactness at O$(\ve^{2})$ originates from the existence of simple relations between Feynman diagrams for the field two-point function and diagrams contributing to the interaction propagator.

For $d_{c}=1$, the O($\ve^{2}$) correction to the scaling exponent $\eta$ is suppressed by a small numerical prefactor. Even in the physical case $D=2$, where the expansion parameter $\ve$ is as large as $\ve = 2$, the extrapolated exponent is dominated by the leading O$(\ve)$ result. In addition, the quantitative values of $\eta$ obtained by direct extrapolation to $D = 2$, $\eta = 0.8$ and $\eta \simeq 0.795$, are close to earlier predictions. The unexpected stability verified in the lowest orders of the $\ve$-expansion and the proximity  to previously reported values suggest that the perturbative analysis of the GCI model could provide an useful framework for quantitatively accurate predictions of the scaling exponent.

\section*{Acknowledgements}
This work was supported by the Netherlands Organisation for Scientific Research (NWO) via the Spinoza Prize.

\appendix
\section{Two-loop self-energy diagram} \label{app:two-loop-self-energy}
Diagrams (c) and (g) in Fig.~\ref{fig:two-loop-feynman} lead to the two-loop integral:
\begin{equation} \label{eq:2L}
a_{D}  = \frac{1}{s^{2}_{D}} \int \frac{{\rm d}^{D}k_{1}}{(2\pi)^{D}} \int \frac{{\rm d}^{D}k_{2}}{(2\pi)^{D}} \frac{\gamma(\hq, \bk_{1}) \gamma(\bk_{1}, \bk_{2}) \gamma(\hq, \bk_{2}) \gamma(\hq-\bk_{1}, \hq-\bk_{2})}{|\bk_{1}|^{4} |\bk_{2}|^{4} |\hq - \bk_{2}|^{4}|\hq - \bk_{1}|^{4} |\bk_{1} - \bk_{2}|^{4}}~,
\end{equation}
where $\gamma(\bk_{1}, \bk_{2}) = k_{1}^{2} k_{2}^{2} - (\bk_{1} \cdot \bk_{2})^{2}$, $\hq = \bq/|\bq|$ is an unit vector and integration runs over dimensionless momenta $\bk_{1}$ and $\bk_{2}$. The diagram presents a logarithmic UV divergence in four dimensions. Because of the finiteness of the $\lambda \bh^{2}$ three-point vertex function, all subdiagrams are finite. The UV divergence is thus of global type: it is generated by the region of integration in which $\bk_{1}$ and $\bk_{2}$ are both simultaneously large. As any global UV divergence, the divergence in $D_{c}(\bq)$ corresponds to a first order pole in dimensional regularization. Following a standard strategy, it is possible to extract the singularity by replacing the integrand in Eq.~\eqref{eq:2L} by any simpler expression which presents the same large-momentum behavior. A convenient choice consists in the replacement $a_{D} \to \bar{a}_D$, with:
\begin{equation} \label{eq:2L-1}
 \bar{a}_{D} = \frac{1}{s_{D}^{2}}\int \frac{{\rm d}^{D}k_{1}}{(2\pi)^{D}} \int \frac{{\rm d}^{D} k_{2}}{(2\pi)^{D}} \frac{\gamma(\hq, \bk_{1}) \gamma(\hq, \bk_{2}) \gamma^{2}(\bk_{1}, \bk_{2})}{\lt(|\bk_{1}|^{4} + \sigma k_{1}^{2}\rt)^{2} \lt(|\bk_{2}|^{4} + \sigma k_{2}^{2}\rt)^{2}\lt|\bk_{1} - \bk_{2}\rt|^{4}}~.
\end{equation}
In this expression $\sigma$ plays the role of an external tension~\cite{guitter_jpf_1989, burmistrov_prb_2018, burmistrov_aop_2018}, modifying the $\bh$ field propagator from $1/k^{4}$ to $1/(k^{4} + \sigma k^{2})$. Imposing a finite $\sigma$ in Eq.~\eqref{eq:2L-1} is necessary in order to avoid infrared divergence of the integral. The dependence of $\bar{a}_{D}$ on the external momentum $\hq$ can now be factorized. The integral takes the form:
\begin{equation}
\bar{a}_{D} = P^{T}_{\alpha \beta}(\hq) P^{T}_{\gamma \delta}(\hq) \int_{\bk_{1}} \int_{\bk_{2}} f(k_{1}^{2}, k^{2}_{2}, \bk_{1} \cdot \bk_{2}) k_{1 \alpha} k_{1 \beta} k_{2 \gamma} k_{2 \delta}~,
\end{equation}
where $P^{T}_{\alpha \beta}(\bq) = \delta_{\alpha \beta} - \hat{q}_{\alpha} \hat{q}_{\beta}$ is the transverse projector and $\int_{\bk} = \int {\rm d}^{D}k/(2\pi)^{D}$ denotes momentum integration. It is then convenient to average over angles~\cite{le-doussal_aop_2018}. By using the relation~\cite{zinn-justin_qft}
\begin{equation}
\int_{\bk_{2}} \,f(k_{1}^{2}, k^{2}_{2}, \bk_{1} \cdot \bk_{2}) k_{2 \gamma} k_{2 \delta} = \int_{\bk_{2}}  \frac{f(k_{1}^{2},  k^{2}_{2}, \bk_{1} \cdot \bk_{2})}{(D-1) k_{1}^{2}} \bigg[\gamma(\bk_{1}, \bk_{2})\delta_{\gamma \delta} +  \lt(D(\bk_{1} \cdot \bk_{2})^{2} - k_{1}^{2}k_{2}^{2}\rt)\frac{k_{1 \gamma} k_{1 \delta}}{k_{1}^{2}}\bigg]
\end{equation}
and the spherical averages~\cite{le-doussal_aop_2018}
\begin{equation}
\begin{gathered} 
\int_{\bk_{1}} g(k_{1}^{2})k_{1\alpha} k_{1\beta} = \frac{\delta_{\alpha \beta}}{D} \int_{\bk_{1}} g(k_{1}^{2}) k_{1}^{2}~,\\
\int_{\bk_{1}} g(k_{1}^{2}) k_{1\alpha} k_{1\beta} k_{1\gamma} k_{1\delta} = \frac{\delta_{\alpha \beta} \delta_{\gamma \delta} + \delta_{\alpha \gamma}\delta_{\beta \delta} + \delta_{\alpha \delta} \delta_{\beta \gamma}}{D(D+2)} \int_{\bk_{1}} g(k_{1}^{2}) k_{1}^{4}~,
\end{gathered}
\end{equation}
we obtain:
\begin{equation}
\begin{split}
\bar{a}_{D} &= \frac{1}{D(D+2)} \int_{\bk_{1}} \int_{\bk_{2}} f(k_{1}^{2}, k_{2}^{2}, \bk_{1} \cdot \bk_{2}) [(D^{2}-1)k_{1}^{2} k_{2}^{2} -2\gamma(\bk_{1}, \bk_{2})]\\
& = \frac{1}{D (D+2) s_{D}^{2}} \int_{\bk_{1}} \int_{\bk_{2}} \frac{\gamma^{2}(\bk_{1}, \bk_{2})[(D^{2}-1)k_{1}^{2} k_{2}^{2} - 2 \gamma(\bk_{1}, \bk_{2})]}{(|\bk_{1}|^{4} + \sigma k_{1}^{2})^{2}(|\bk_{2}|^{4} + \sigma k_{2}^{2})^{2}|\bk_{1} - \bk_{2}|^{4}}~.
\end{split}
\end{equation}
By introducing integration over five Schwinger-type parameters $t_{i}$ ($i = 1, .., 5$), the expression for $\bar{a}_{D}$ can be represented as
\begin{equation} \label{eq:schwinger}
\begin{gathered}
\bar{a}_{D} = \frac{1}{D(D+2)s_{D}^{2}}\int_{0}^{\infty} {\rm d}t_{1} {\rm d}t_{2} {\rm d}t_{3} {\rm d}t_{4} {\rm d}t_{5} \int_{\bk_{1}} \int_{\bk_{2}} \bigg\{ \bigg(\prod_{i=1}^{5}t_{i}\bigg)  \gamma^{2}(\bk_{1}, \bk_{2}) [(D^{2}-1)k_{1}^{2} k_{2}^{2} - 2 \gamma(\bk_{1}, \bk_{2})]\\
\times \exp \big[-\sum_{a, b = 1}^{2}M_{ab} (\bk_{a} \cdot \bk_{b}) - (t_{2} + t_{4})\sigma\big]\bigg\}~,
\end{gathered}
\end{equation}
where $M$ is the $2 \times 2$ matrix:
\begin{equation} \label{def:M}
M = \begin{bmatrix}
     t_{1} + t_{2} + t_{5} & - t_{5}\\
     -t_{5} & t_{3} + t_{4} + t_{5}
    \end{bmatrix}
\end{equation}
and integrals over all five variables $t_{i}$ ($i = 1, .., 5$) run from $0$ to $\infty$. Momentum integrals in Eq.~\eqref{eq:schwinger} are determined by moments of a Gaussian distribution. It is convenient to express moments by differentiation with respects to the matrix elements $M_{ab}$. This leads to the representation
\begin{equation}
\begin{gathered}
\bar{a}_{D}  = \frac{1}{D(D+2) s_{D}^{2}}\int_{0}^{\infty}{\rm d}t_{1} {\rm d}t_{2} {\rm d}t_{3} {\rm d}t_{4} {\rm d}t_{5} \int_{\bk_{1}} \int_{\bk_{2}}\bigg\{\bigg( \prod_{i=1}^{5}t_{i}\bigg) {\rm e}^{-(t_{2} + t_{4})\sigma}\\
\times \bigg[(D^{2}-1)\frac{\pa^{2}}{\pa M_{11}\pa M_{22}} - 2 \bigg(\frac{\pa^{2}}{\pa M_{11} \pa M_{22}} - \frac{1}{4} \frac{\pa^{2}}{\pa M_{12}^{2}}\bigg)\bigg] \bigg[\frac{\pa^{2}}{\pa M_{11}\pa M_{22}} - \frac{1}{4} \frac{\pa^{2}}{\pa M_{12}^{2}}\bigg]^{2} {\rm e}^{-\sum_{a, b} M_{ab}(\bk_{a} \cdot \bk_{b})}\bigg\}~,
\end{gathered}
\end{equation}
where $M_{12} = M_{21}$ is considered as a single independent variable. By using the relation
\begin{equation}
\bigg[\frac{\pa^{2}}{\pa M_{11} \pa M_{22}} - \frac{1}{4} \frac{\pa^{2}}{\pa M_{12}^{2}} \bigg] (\det M)^{-\omega} = \frac{\omega (2\omega -1)}{2}(\det M)^{-\omega-1}
\end{equation}
after momentum integration, we obtain:
\begin{equation}
\begin{gathered}
\bar{a}_{D} = \frac{(D^{2}-1) (D+4)}{64 (4\pi)^{D}s_{D}^{2}} \int_{0}^{\infty} {\rm d}t_{1} {\rm d}t_{2} {\rm d}t_{3} {\rm d}t_{4} {\rm d}t_{5} \,\bigg\{\bigg(\prod_{i=1}^{5} t_{i}\bigg) \frac{{\rm e}^{-(t_{2} + t_{4})\sigma} }{(\det M)^{\frac{D+6}{2}}} \\
\times \Big[ (D^{2}-1)(D+6) \frac{M_{12}^{2}}{\det M} +   (D^{2}-1)(D+4) - 2(D+3)  \Big]\bigg\}~.
\end{gathered}
\end{equation}
Near $D = 4$ regular functions of $D$ can be replaced by their four dimensional value. Using the explicit form of the matrix $M$, Eq.~\eqref{def:M}, and integrating over $t_{5}$ gives:
\begin{equation}
\begin{gathered}
\bar{a}_{D} \approx 4 \int_{0}^{\infty} {\rm d}t_{1} {\rm d}t_{2} {\rm d}t_{3} {\rm d}t_{4} \bigg\{\bigg(\prod_{i=1}^{4} t_{i}\bigg) \bigg(9 + \frac{53}{5} \frac{s^{2}}{(t_{1} + t_{2}) (t_{3} + t_{4})}\bigg)\frac{{\rm e}^{-(t_{2} + t_{4})\sigma}}{s^{4}(t_{1} + t_{2})^{D/2} (t_{3} + t_{4})^{D/2}} \bigg\}~,
\end{gathered}
\end{equation}
where $s = t_{1} + t_{2} + t_{3} + t_{4}$. By the change of variables
\begin{equation}
t_{1} = (1-x_{1})y s~,\qquad t_{2} = x_{1} ys ~,\qquad t_{3} = (1-x_{2})(1-y)s~,\qquad t_{4} = x_{2} (1-y)s~,
\end{equation}
the integral can be rewritten in a Feynman-type parametrization:
\begin{equation}
\begin{gathered}
\bar{a}_{D} \approx 4 \int_{0}^{1} {\rm d}x_{1} \int_{0}^{1} {\rm d}x_{2} \int_{0}^{1} {\rm d}y \int_{0}^{\infty} {\rm d}s \;\big\{ x_{1}(1-x_{1})x_{2}(1-x_{2}) y^{3- \frac{D}{2}} (1-y)^{3-\frac{D}{2}}s^{3-D} \Big(9 + \frac{53}{5 y (1-y)}\Big)\\
\times \exp \big[-(x_{1} y + x_{2}(1-y))s\sigma\big] \big\}~.
\end{gathered}
\end{equation}
Integration over $s$ generates a first order pole at $D = 4$. The remaining integrals, being finite, can be calculated by replacing $D = 4$ in the integrand function. The result is:
\begin{equation}
\begin{gathered}
\bar{a}_{D} \approx 4 \Gamma(4-D)\int_{0}^{1} {\rm d}x_{1} \int_{0}^{1} {\rm d}x_{2} \int_{0}^{1} {\rm d}y\; \bigg[x_{1}(1-x_{1})x_{2}(1-x_{2})\bigg(9  y(1-y) + \frac{53}{5}\bigg) \bigg]\approx \frac{121}{90(4 - D)}~.
\end{gathered}
\end{equation}
The two-loop integral $a_{D}$, therefore, behaves near four dimensions as:
\begin{equation}
a_{D} = \frac{121}{90 \ve} + {\rm O}(1)~.
\end{equation}



\end{document}